\documentclass[a4paper,11pt]{article}
\pdfoutput=1 

\usepackage{jcappub} 
\usepackage[T1]{fontenc} 
\usepackage{bm}
\usepackage{lineno}
\usepackage{doi}
\usepackage{appendix}

\title{\boldmath Recovering CMB polarization maps with neural networks: Performance in  realistic simulations}

\author[a,b,c,d,1]{J. M. Casas,\note{Corresponding author.}}
\author[c,d]{L. Bonavera,}
\author[c,d]{J. González-Nuevo,}
\author[e,f]{G. Puglisi,}
\author[g,h]{C. Baccigalupi,}
\author[c,d]{S. R. Cabo,}
\author[g,h]{M. M. Cueli,}
\author[c,d]{D. Crespo,}
\author[i,d]{C. González-Gutiérrez,}
\author[j,d]{F.J. de Cos}

\affiliation[a]{Instituto de Astrofísica de Canarias, E-38205 La Laguna, Tenerife, Spain}
\affiliation[b]{Universidad de La Laguna, Departamento de Astrofísica, E-38206 La Laguna, Tenerife, Spain}
\affiliation[c]{Departamento de F{\'i}sica, Universidad de Oviedo,\\C. Federico Garc{\'i}a Lorca 18, 33007 Oviedo, Spain}
\affiliation[d]{Instituto Universitario de Ciencias y Tecnolog{\'i}as Espaciales de Asturias (ICTEA),\\C. Independencia 13, 33004 Oviedo, Spain}
\affiliation[e]{Dipartimento di Fisica e Astronomia, Universitá degli Studi di Catania,\\Via S. Sofia, 64, 95123, Catania, Italy}
\affiliation[f]{INFN – Sezione di Catania,\\Via S. Sofia 64, 95123 Catania, Italy}
\affiliation[g]{SISSA,\\Via Bonomea 265, 34136 Trieste, Italy}
\affiliation[h]{INFN-Sezione di Trieste,\\Via Valerio 2, 34127 Trieste,  Italy}
\affiliation[i]{Departamento de Inform{\'a}tica, Universidad de Oviedo,\\Edificio Departamental 1. Campus de Viesques s/n, E-33204, Gij{\'o}n, Spain}
\affiliation[j]{Escuela de Ingeniería de Minas, Energía y Materiales, Universidad de Oviedo,\\C. Independencia 13, 33004 Oviedo, Spain}

\emailAdd{casasjm@uniovi.es}

\abstract{
Recovering the polarized cosmic microwave background (CMB) is essential for shedding light on the exponential expansion of the very early Universe, known as cosmic inflation. Achieving this goal requires not only improved instrumental sensitivity but also the development of robust and diverse data analysis techniques. In this work, we explore a novel component separation approach based on neural networks, previously validated using realistic \textit{Planck} temperature simulations, to reconstruct the Stokes $Q$ and $U$ polarization maps.

To validate the method, we first test the network on realistic \textit{Planck} sky simulations of regions deliberately excluded from the training set. We compare the input and output $EE$ and $BB$ power spectra, finding a mean absolute error of $0.1 \pm 0.3~\mu K^{2}$ for the $E$-mode and $-0.1 \pm 0.3~\mu K^{2}$ for the $B$-mode. These results demonstrate a partial recovery of the $E$-mode and a limited recovery of the $B$-mode, the latter remaining dominated by residual \textit{Planck} noise. We then apply the trained network to public \textit{Planck} observations, recovering CMB polarization maps broadly consistent with those obtained using the Commander method. The recovered $EE$ spectra differ by less than 5\% from the reference at intermediate and small angular scales, although significant discrepancies remain at large scales, which may impact cosmological interpretations. These results, while encouraging, clearly reflect the limitations of the current setup and motivate further improvements in training data and methodology.

Based on these findings, we conclude that neural network-based methods show potential as component separation techniques in polarization CMB experiments. However, substantial improvements and more comprehensive analyses are necessary before these methods can provide reliable high-precision cosmological estimates.
}

\begin{document}
\maketitle
\flushbottom

\section{Introduction}
\label{sec:introduction}

The cosmic microwave background (CMB) has, over the past few decades, greatly aided the scientific community in understanding the nature and evolution of the Universe. It originates from an epoch approximately 380000 years after the Big Bang, when photons decoupled from baryons in an event known as recombination \citep{WEI08}. Although isotropic to first order, the CMB exhibits small fluctuations in temperature and polarization, known as anisotropies. These can be described by approximating the celestial sphere using the Stokes $Q$ and $U$ parameters \cite{HU97}; specifically, its polarization can be characterized by combining these parameters into gradient and curl modes, referred to as $E$ and $B$ modes, respectively \citep{ZAL97}. The $E$ modes are associated with scalar (density) perturbations in the early Universe, whereas the $B$ modes arise from tensor (gravitational) perturbations, produced by gravitational lensing of the CMB by intervening large-scale structures as well as by inflationary gravitational waves.

In recent decades, various experiments have mapped the CMB with high precision. In particular, the Planck mission \citep{PLA_18_I} observed the microwave sky over the past decade and produced unprecedentedly detailed maps, which have significantly constrained the fundamental properties of the Universe and characterized the main microwave emissions from our Galaxy, especially in temperature maps. Current and future missions, such as the Simons Observatory \citep{SO}, LiteBIRD \citep{LITEBIRD}, and PICO \cite{Han19}, aim to improve constraints on the CMB polarization, in particular the $B$-mode power spectrum at large angular scales, in order to search for primordial gravitational waves in the cosmological signal.

With upcoming experiments more sensitive than \textit{Planck}, the main challenge will be to accurately characterize the microwave sky, as the CMB polarization is contaminated at all sky regions and scales by Galactic and extragalactic emissions. This challenge is particularly critical for constraining the tensor-to-scalar ratio of $B$- and $E$-modes down to $r = 0.001$ (\cite{KRA16}, \cite{PUG18}). A precise understanding of the polarization properties of these emissions, known as foregrounds, is essential for effectively separating the CMB signal from other sources, a process referred to as component separation. Although several physical models are continually being tested and improved to address the challenges of the coming decade (the main approaches are described in detail in \cite{FUS23}), machine learning models, such as neural networks, have only recently begun to be explored for this application, despite having demonstrated promising results in other areas of astronomy \citep{SMI23}. Neural networks are promising tools for such analyses due to their ability to learn complex non-linear behaviors directly from the data \citep{GOO16}, which may help disentangle the diverse physical processes contributing to the microwave sky.

To date, only a few studies have demonstrated the potential of neural networks for various tasks in CMB research, particularly in component separation. For instance, they have first been validated using temperature maps (\cite{PET20}, \cite{WAN22}, \cite{CAS22a}, \cite{CAS22b}), and have subsequently shown encouraging results in polarization analysis \cite{PEN23}. Despite the well-known complexity of the polarized sky, neural networks have recently begun to be accurately tested with real observations \cite{CAS23}, suggesting that they could serve as a valuable complement to traditional methods in future CMB experiments. Another relevant applications in the recent years are for primordial non-Gaussianity \cite{NOV15}, delensing \cite{CAL19}, cosmic string detection \cite{CIU19}, foreground characterization (\cite{FAR20}, \cite{KRA21}) and inpainting \cite{MON21}. Neural networks, however, are still viewed with skepticism by parts of the community due to their strong dependence on training model data, their lack of physical interpretability, and some initial controversial results \cite{NOR18}.

In this work, we aim to assess the performance of a neural network similar to that used in our previous study on temperature maps \cite{CAS22b} in recovering the CMB polarization signal. This paper is organized as follows: Section~\ref{sec:data} describes the data used in this study; Section~\ref{sec:methodology} outlines the adopted methodology; Section~\ref{sec:results} presents our results on both simulations and real observations; and finally, Section~\ref{sec:conclusions} summarizes our conclusions.

\section{Data}
\label{sec:data}

\subsection{\textit{Planck} observations}
\label{sec:observations}

The Planck mission, operated by the European Space Agency (ESA), was designed to map the full-sky cosmic microwave background (CMB) anisotropies with high sensitivity and angular resolution. Launched in 2009 and operational until 2013, \textit{Planck} carried two instruments: the Low Frequency Instrument (LFI), covering 30, 44, and 70 GHz, and the High Frequency Instrument (HFI), covering 100-857 GHz. The mission achieved full-sky coverage in both temperature and polarization across nine frequency channels, enabling robust foreground modeling and component separation.

The final legacy data release, known as Planck 2018 or Planck Release 3 (PR3), comprises a reprocessing of the entire mission dataset, incorporating improved calibration, beam characterization, and systematic error mitigation, particularly in the HFI polarization channels. This release provides temperature and polarization power spectra (TT, TE, EE) with cosmic variance-limited precision up to multipoles $l \sim 1500$, as well as robust measurements of the lensing potential power spectrum derived from four-point correlations.

The associated Planck 2018 likelihoods were constructed using advanced component separation algorithms (Commander, NILC, SEVEM, and SMICA\footnote{These methods are described in detail in \cite{PLA_2015_IX}.}) and validated through extensive simulations. The precision of these measurements provided strong support for the minimal six-parameter $\Lambda$CDM model, leaving limited room for deviations from standard inflationary predictions or for the presence of additional relativistic species.

\subsection{Simulations}
\label{sec:simulations}

To train and validate the neural network before testing its performance on real observations, we use a set of realistic microwave sky simulations representing the sky as observed by the HFI Planck instrument at 100, 143, and 217 GHz. These channels are in fact the least affected by noise in polarization, with levels of 1.96, 1.17, and 1.75 $\mu K$deg, respectively \citep{PLA_18_I}.

The simulations include a lensed CMB map together with Galactic interstellar dust and synchrotron emission maps. These foreground maps were downloaded from the Planck Legacy Archive (PLA\footnote{https://pla.esac.esa.int}), which incorporates updated foreground models from PR3 \citep{PLA_18_IV}. To test the robustness of our method, we also use alternative foreground models, different from those used for training, by employing the Python Sky Model (PySM; \cite{THO17}, \cite{ZON21}, \cite{Panexp_2025}).

Our method is based on convolutional neural networks (CNNs). Since similar approaches designed for spherical data exist \cite{KRA19}, but CNNs generally operate on flat grids, we must project spherical maps onto two-dimensional planes. Initially, we observed significant $E$-to-$B$ leakage when using the gnomonic projection from Healpy for almost all patch sizes. To address this, we adopt the projection method described in \citep{KRA21}, since we found that it also suffered $E$-to-$B$ leakage but only for big sky patches.

Following this approach, we divide the microwave sky maps into square patches of 256$\times$256 pixels, with a pixel size of 90 arcseconds, and verify that no $E$-to-$B$ leakage occurs by comparing the average input signal from all sky patches to the theoretical prediction. Additionally, we inject extragalactic radio sources into these patches using the model from \cite{TUC11} and the CORRSKY software \citep{GN05}. Finally, instrumental noise at \textit{Planck} levels is then added to each simulation, providing a different noise realization for each patch and minimizing redundancy in the training data. The noise is modeled as isotropic white noise with the RMS levels defined above. Although being a simplification with respect to a scan path modulated white noise and/or an end-to-end simulation, it allows to have a first approximation of the performance of this kind of network architectures on polarization data, especially at the arcmin and sub-arcmin scales.

For each position, we generate three frequency patches, referred to as the Input Total, which include all emissions plus noise at 100, 143, and 217 GHz. We also retain the corresponding CMB-only patch at 143 GHz as the label. The training set comprises 10000 simulations in the northern sky region ($90^{\circ} < b < -60^{\circ}$), where $b$ denotes Galactic latitude, including the Galactic plane and other complex, physically distinct regions. For each test, the validation set consists of 1000 simulations located in the southern sky region ($-60^{\circ} < b < -90^{\circ}$), encompassing regions entirely different from those in the training set. This setup allows us to test the robustness and generalization capability of our method.

\begin{figure*}
\centering
\includegraphics[width=15cm, height=3.2cm]{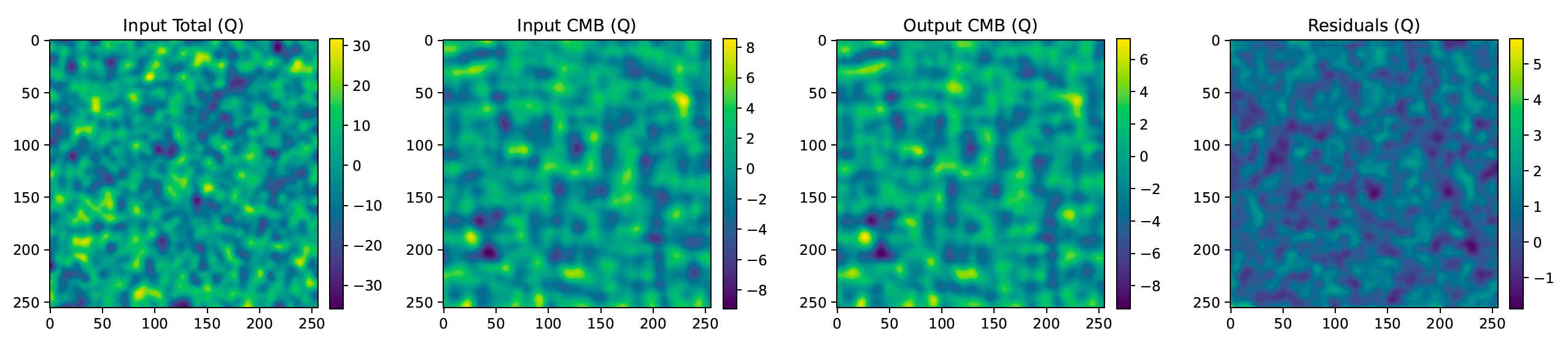}
\includegraphics[width=15cm, height=3.2cm]{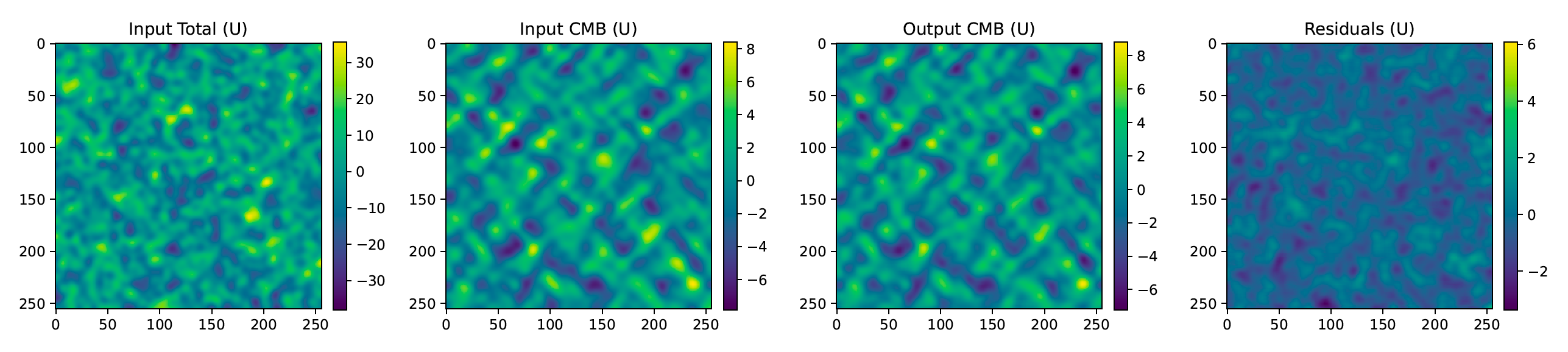}
\caption{An example of a realistic simulation used in this work. From left to right, the columns display the Input Total patch (with all simulated emissions), the input CMB (which is also part of the Input Total), the neural network output, and the residual patch, defined as the difference between input and output. The colorbar units in all panels are in $\mu K$.}
\label{Fig:Patches}
\end{figure*}

An example of simulated patches at a random sky position is shown in Figure~\ref{Fig:Patches}. In this figure, the first column from the left shows the Input Total patch containing all polarized emissions; the second column shows the CMB contribution within the Input Total. The top and bottom rows correspond to the Stokes $Q$ and $U$ maps, respectively.

\section{Methodology}
\label{sec:methodology}

An artificial neural network is a type of machine learning model composed of computational units called neurons, which are organized into layers \cite{GOO16}. Each neuron contains weights and biases, parameters that are iteratively updated during training. When input data pass through the network layers, a loss function quantifies the difference between the network’s output and the true target. This loss is minimized using the backpropagation algorithm \cite{Rum86}, which computes gradients that update the weights, biases, and filters throughout the network. By repeating this process over many training epochs, the network progressively improves its ability to generalize to new data.

Neural networks are especially effective for analyzing image data, as they apply convolutions and deconvolutions to learn spatial features from input images. For instance, they have been widely and successfully applied to image segmentation tasks in recent years \citep{LON15}. In this context, the learnable weights are organized into matrices called kernels, which are optimized during training to extract meaningful features at multiple scales.

In this work, we extend the Cosmic Microwave Background Extraction Neural Network (CENN) first presented in \citep{CAS22b} to evaluate its performance on CMB polarization maps. CENN is a fully convolutional neural network based on the U-Net architecture \citep{RON15}, designed to produce clean CMB maps from noisy sky patches by performing a sequence of convolutions and deconvolutions. The network minimizes a loss function that measures the difference between the predicted and true CMB maps. The mathematical and physical principles underlying this approach are detailed in \citep{GOO10} and \citep{CAS22b}; here, we briefly summarize the architecture and its operation.

The network processes each Input Total patch by successively convolving the input with learnable filters across six consecutive convolutional blocks. Each block extracts hierarchical features, while an activation function introduces non-linearity. The encoded information is then upsampled through six corresponding deconvolutional blocks, each linked to its convolutional counterpart via skip connections. This structure allows the network to combine global context with fine details, distinguishing the CMB signal from contaminating foregrounds. In the final block, the feature maps are merged into a single output patch representing the recovered CMB. The mean squared error loss function is computed between this output and the true CMB map, providing the gradient used to update all weights and filters via backpropagation and the AdaGrad optimizer. Training in this work is particularly performed over 1000 epochs.

\begin{figure*}[ht]
\centering
\includegraphics[width=15cm]{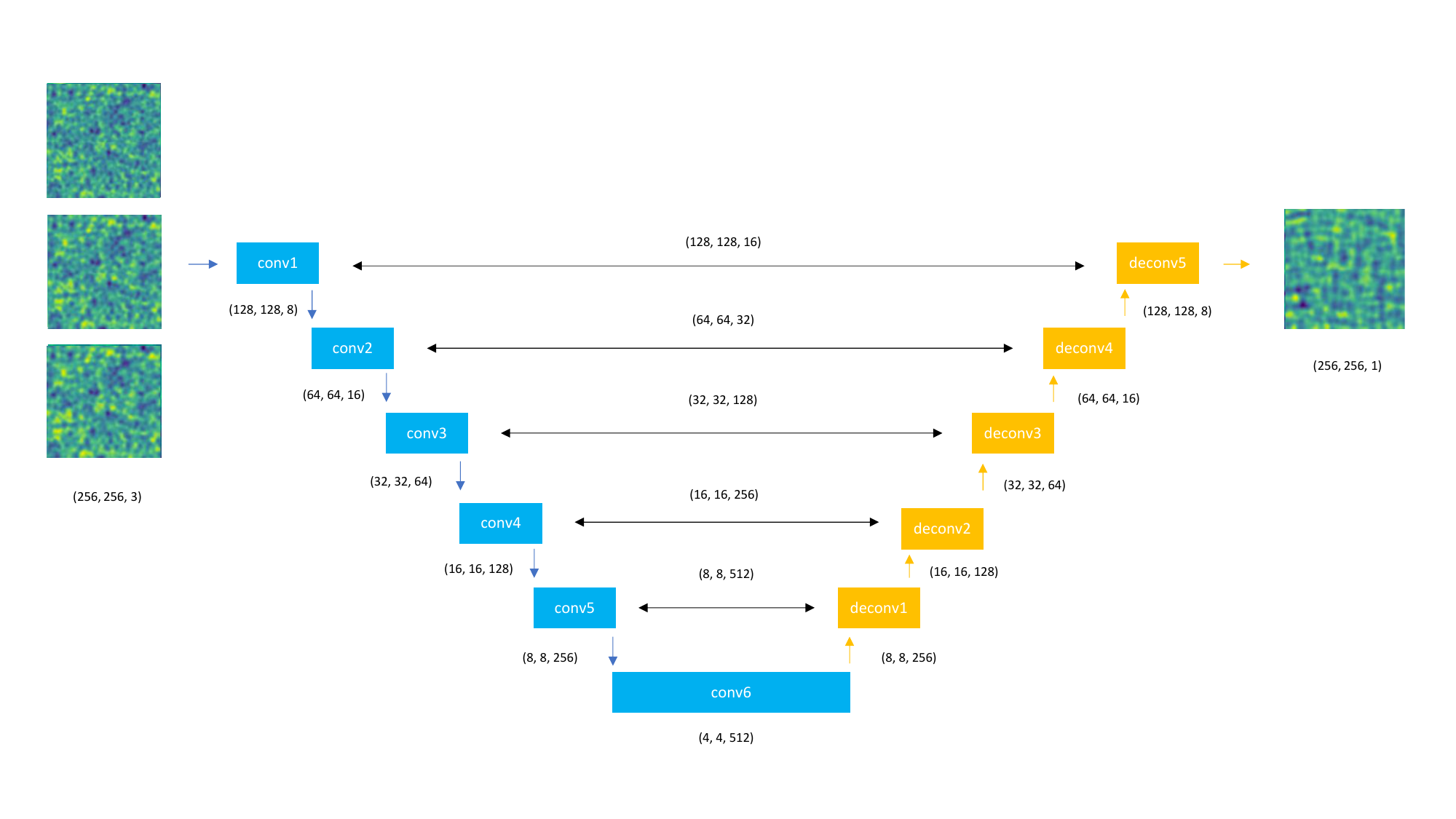}
\caption{Architecture of CENN, the neural network used in this work. We show for visualization purposes the kind of input and output patches the network reads and outputs, respectively.}
\label{Fig:Architecture}
\end{figure*}

The specific architecture is illustrated in Figure~\ref{Fig:Architecture} and can be described as follows. The encoder consists of six convolutional blocks, each containing convolutional and pooling layers with kernel sizes of 9, 9, 7, 7, 5, and 3, respectively, and with 8, 2, 4, 2, 2, and 2 kernels per block. The corresponding numbers of filters are 8, 16, 64, 128, 256, and 512, and a subsampling factor of 2 is applied to downsample the feature maps. All layers use padding type “same” to preserve spatial dimensions and a leaky ReLU activation function for characterizing the quantity of non-linear information allowed to pass through the next layer.

The decoder comprises six deconvolutional blocks, each with deconvolutional and pooling layers using 2, 2, 2, 4, 2, and 8 kernels of sizes 3, 5, 7, 7, 9, and 9, respectively. The numbers of filters decrease symmetrically to 256, 128, 64, 16, 8, and finally 1 for the output map. As in the encoder, all layers apply “same” padding and a leaky ReLU activation function.

\section{Results}
\label{sec:results}

This section presents the results of validating our network in three stages: first with ideal simulations without instrumental noise, then with realistic \textit{Planck} simulations, and finally with real PR3 observations. Additional tests using realistic simulations are performed to assess the robustness of the methodology. The evaluation primarily involves comparing input and output maps, computing residuals, and estimating the power spectra for both the $E$- and $B$-mode signals.

\subsection{Validation by simulations: ideal case}
\label{sec:results_ideal_case}

\begin{figure*}[t]
\centering
\minipage{0.5\textwidth}
\includegraphics[width=\linewidth]{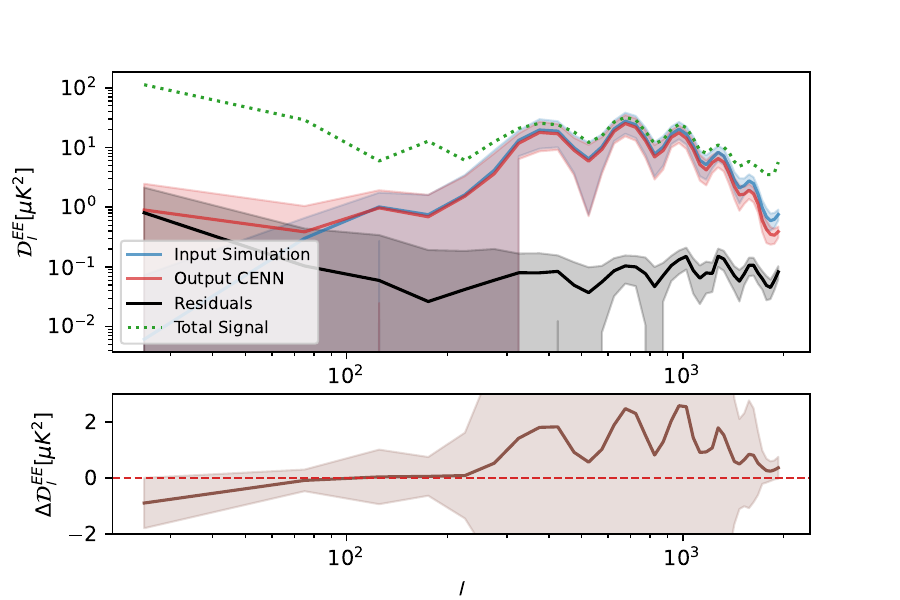}
\endminipage\hfill
\minipage{0.5\textwidth}%
  \includegraphics[width=\linewidth]{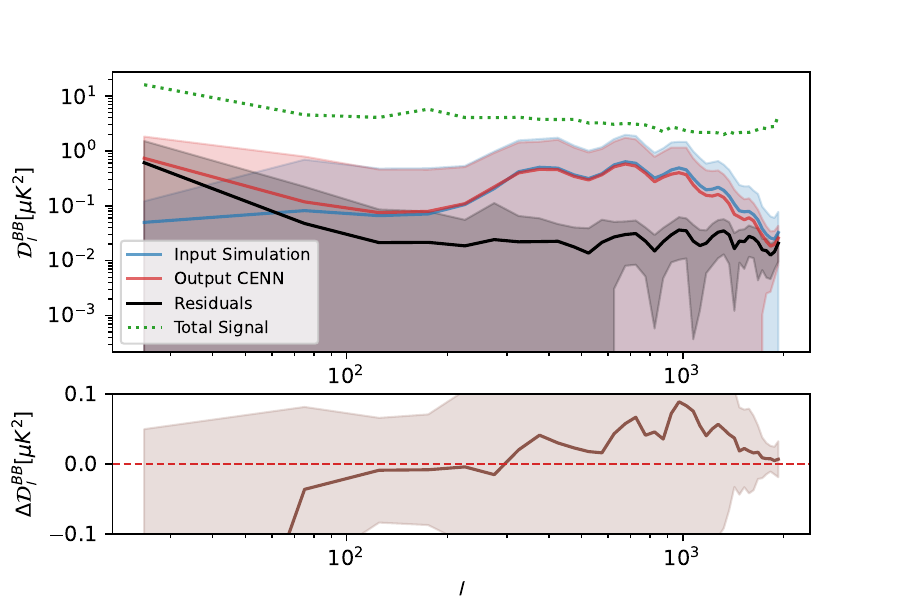}
\endminipage
\caption{Recovered $E$- and $B$-mode power spectra using CENN for the ideal case without instrumental noise. Top left: input $E$-mode spectrum (blue), network's output (red), and residuals (black); the green dotted line shows the average Input Total spectrum at 143 GHz. The bottom left panel shows the absolute difference between input and output spectra. Right panels show the same for the $B$-mode. Shaded regions indicate the $1\sigma$ standard deviation per bin.}
\label{Fig:Ideal_Case}
\end{figure*}

We begin by training and validating the network in an idealized scenario, using simulations that include only the CMB and astrophysical foregrounds, with no instrumental noise. As described in Section~\ref{sec:simulations}, to ensure robust testing, the network is trained on sky patches within the region $90^{\circ} < b < -60^{\circ}$, where $b$ is the Galactic latitude, while the remaining sky regions are reserved for validation. In total, we generate 1000 $Q$ and $U$ polarization patches for validation and compute the average $EE$ and $BB$ power spectra of the input, output and residual maps using NaMaster \citep{NAMASTER}. Figure~\ref{Fig:Ideal_Case} shows the resulting spectra: the input CMB is shown in blue, the network's output in red, and the residuals in black. Shaded regions indicate the $1\sigma$ standard deviation for each multipole bin. For reference, the green dotted line shows the average spectra of the Input Total at 143 GHz, representing the total signal plus foreground contamination provided to the network during validation\footnote{Only the total contribution at 143 GHz is shown, as this is the frequency at which the recovered CMB is produced.}.

For multipoles $l > 200$, we found a mean absolute error of $0.5 \pm 1~\mu K^{2}$ for the $E$-mode and $0.05 \pm 0.1~\mu K^{2}$ for the $B$-mode, corresponding to relative errors below 5\% for $E$ and about 10\% for $B$. The residuals remain well below the input signal: their mean is typically between $10^{-2}$ and $10^{-1}~\mu K^{2}$ for $E$, and around $10^{-2}~\mu K^{2}$ for $B$, implying that residuals are roughly two orders of magnitude smaller than the input $E$ signal, and about an order of magnitude smaller than the $B$ signal. Furthermore, the network seems to be able to recover the $B$-mode spectra from the maps despite the presence of foregrounds, highlighting the possibility of using this kind of methods in future high-sensitivity CMB experiments targeting this weak signal. Nevertheless, both input and output spectra exhibit relatively large uncertainties, which can be reduced with further improvements in training and network design.

Due to the limited patch size of approximately 6.4 square degrees, the statistical power at large angular scales is restricted. Consequently, our analysis indicates an excess in the recovered $B$-mode signal for $l < 200$. While increasing the patch size could improve performance at large scales, preliminary tests have revealed significant $E$-to-$B$ leakage when projecting larger sky regions onto a flat plane, a technical challenge that warrants dedicated mitigation in future work.

Overall, these results show that a relatively simple fully convolutional architecture such as CENN can perform component separation on simulated polarized CMB maps in sky regions not seen during training. This motivates the use of machine learning-based methods for foreground removal in polarization data and encourages further development of such approaches for application in next-generation CMB experiments with greater sensitivity than \textit{Planck}. However, when extending this architecture to the sphere, the level of uncertainty introduced by this kind of networks, particularly at large angular scales, must be thoroughly analyzed, especially if the goal is to extract the CMB map for use in sub-percent precision cosmological analyses. 

\subsection{Validation by simulations: realistic case}
\label{sec:results_realistic_case}

Building upon the results described in the previous section, we extended our analysis to a more realistic scenario by incorporating instrumental isotropic white noise at \textit{Planck}-level sensitivities (1.96, 1.17, and 1.75 $\mu K\text{deg}$ for the 100, 145, and 217 GHz channels, respectively; see \citep{PLA_18_I}). The neural network was then retrained using these more realistic simulations under the same sky conditions as before: training was performed on the northern region ($90^{\circ} < b < -60^{\circ}$), while validation was conducted on the southern region ($-60^{\circ} < b < -90^{\circ}$). To enhance the signal-to-noise ratio prior to training, we smoothed the input data with a 15' FWHM Gaussian kernel, a standard practice in CMB component separation studies (\cite{PLA_15_IX}; \cite{PLA_15_X} and references therein). This smoothing, however, limits the analysis to multipoles up to $l \sim 1000$.

An example of the resulting $Q$ and $U$ maps generated by CENN is shown in Figure~\ref{Fig:Patches} (third column). The fourth column illustrates the residual patches, obtained as the difference between input and output maps. For this analysis, all maps were smoothed with a 15' FWHM. We then computed the power spectra for the input, output and residual maps using NaMaster.

\begin{figure*}[t]
\centering
\minipage{0.5\textwidth}
\includegraphics[width=\linewidth]{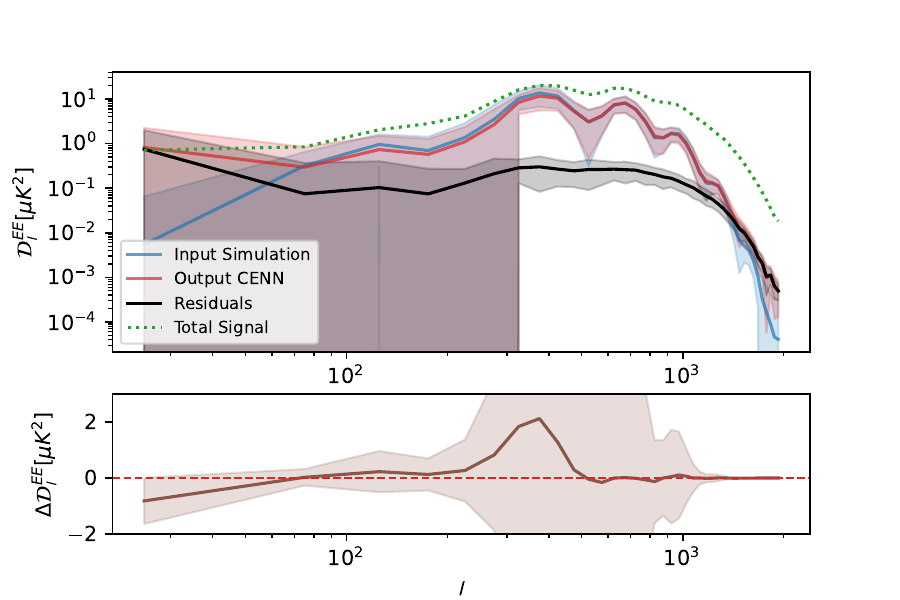}
\endminipage\hfill
\minipage{0.5\textwidth}%
  \includegraphics[width=\linewidth]{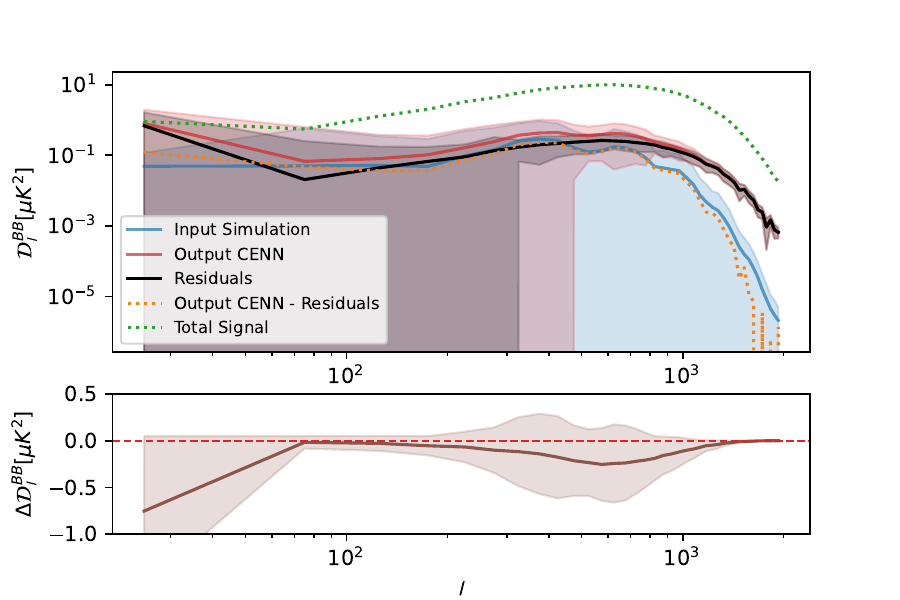}
\endminipage
\caption{Recovered $E$- and $B$-mode power spectra for the realistic case with Planck-level instrumental noise and 15' FWHM smoothing. Top left panel: comparison between the input $E$-mode spectrum from the simulations (blue) and the CENN output (red); the residuals (difference between input and output) are shown in black. The green dotted line indicates the average spectrum of the Input Total signal at 143 GHz. For reference, the orange dotted line shows the difference between the CENN output and the residuals, approximating the estimated recovered CMB signal. The bottom left panel displays the absolute difference between input and output spectra. The right panel presents the same analysis for the $B$-mode. Shaded areas represent the $1\sigma$ standard deviation per multipole bin.}
\label{Fig:15arcmin_case}
\end{figure*}

Figure~\ref{Fig:15arcmin_case} summarizes the network’s performance for this realistic case, with color coding consistent with the previous analysis. For the $E$-mode, we find a mean absolute error of approximately $0.1 \pm 0.3~\mu K^{2}$ and mean residuals of about $0.1~\mu K^{2}$. For the $B$-mode, the mean absolute error is $-0.1 \pm 0.3~\mu K^{2}$, with residuals at comparable levels. These results are broadly similar to the noise-free case but exhibit slightly higher residuals, particularly in the $BB$ spectrum, as well as an underestimation of the $B$-mode instead of the overestimation seen in the previous case. Since the foreground contamination remains consistent with the ideal case, the increase in the uncertainty is primarily attributable to the addition of instrumental noise. The residual structures in the output likely reflect noise features that resemble the CMB signal after smoothing, but that the network cannot fully disentangle. This finding has important implications for future experiments: with lower instrumental noise than \textit{Planck}, since the performance of CENN could approach the near-ideal recovery demonstrated in the noise-free scenario, highlighting its use for next-generation CMB polarization surveys. However, it should be noted that the estimations, when compared with the noise-free case, indicate that future training of this type of neural network method should incorporate more complex noise models than the one used in this work. While the current setup provides a useful first approximation of performance with such data maps, real observations and precision cosmology will require significant improvements in the quality of the training data.

Importantly, residual levels at large angular scales remain comparable to those in the idealized case, indicating that the inclusion of instrumental noise does not substantially degrade CENN’s capacity to mitigate foregrounds at these scales. As discussed in Section~\ref{sec:results_ideal_case}, the limitations at low multipoles appear more strongly linked to the finite patch size and the intrinsic complexity of foreground separation, especially for diffuse dust, than to instrumental noise.

Nevertheless, as shown in Figure~\ref{Fig:15arcmin_case}, the $B$-mode remains more sensitive to both noise and residual foregrounds, as expected. The orange curve, representing the difference between the network output and the residuals, closely tracks the recovered CMB signal. Its agreement with the output spectrum indicates that the network tends to preserve the underlying CMB component, even when the true signal is subdominant to noise. This behavior suggests a degree of robustness, where the network favors minimizing distortions of the CMB signal, albeit at the expense of leaving some residual noise unremoved. 

To explore whether further refinement of the output was feasible, we tested the use of a secondary neural network with an identical architecture with respect CENN, trained to map the initial network’s outputs to the original noise-free CMB signal. However, this additional step did not yield significant improvements, indicating that the current results likely reflect the statistical limits imposed by the low signal-to-noise ratio, particularly for the $B$-mode. Therefore, achieving more accurate component separation for such data will likely require either more sophisticated machine learning architectures or higher signal-to-noise measurements.

For context, a first-order comparison with the traditional component separation methods employed by Planck (SMICA, NILC, Commander, SEVEM) shows that for the $E$-mode, CENN achieves similar performance: Planck reported mean absolute errors of about $1 \pm 6~\mu K^{2}$ for comparable simulations smoothed to 10' FWHM \citep{PLA_2015_IX}, with reliable recovery up to $l \sim 700$, beyond which noise dominates. For the $B$-mode, however, conventional methods did not attempt a direct estimation nor provided an upper limit in comparable setups, making direct comparisons infeasible for this signal component.

\begin{figure}[ht]
\centering
\includegraphics[width=0.7\linewidth]{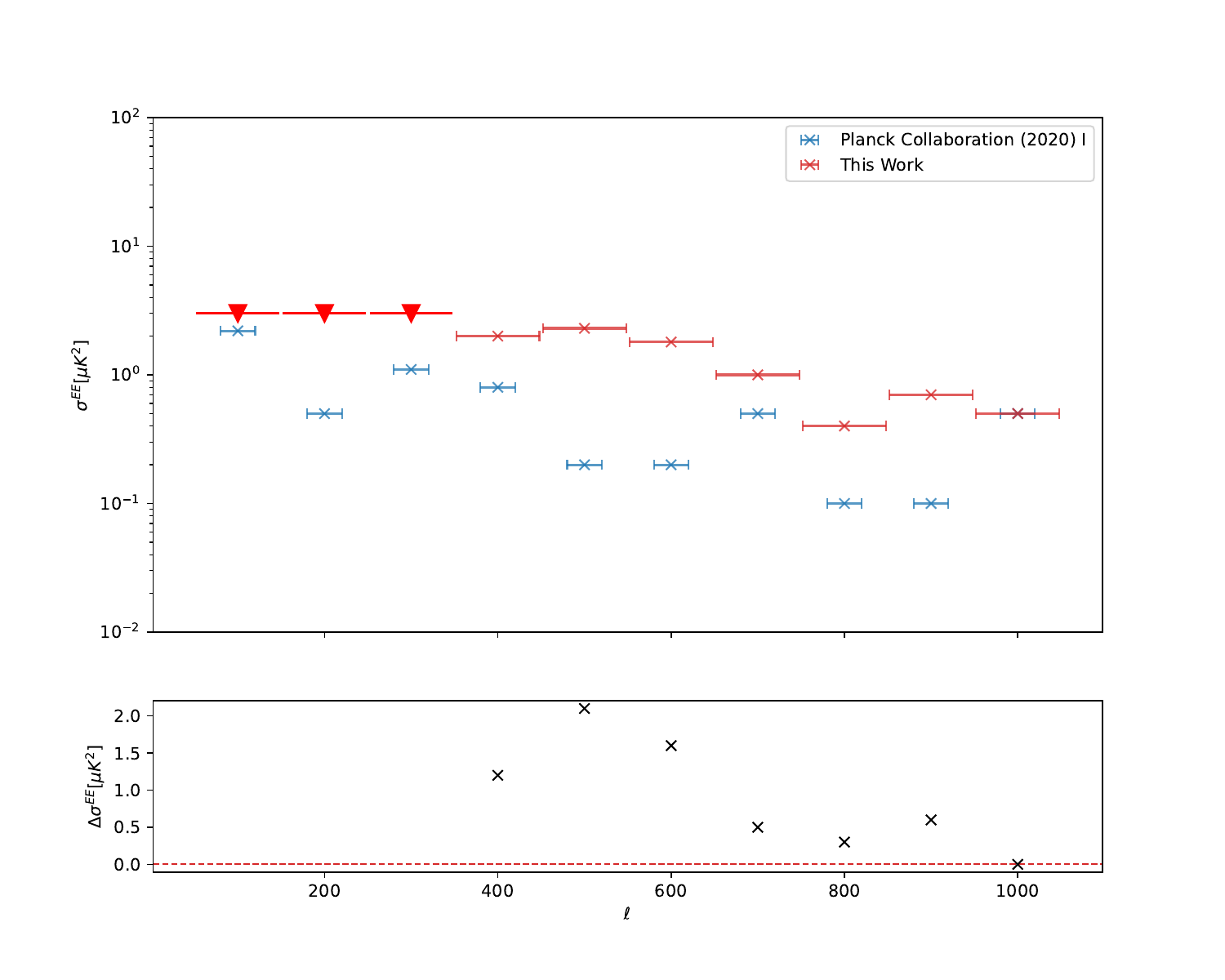}
\caption{Comparison between the CENN systematic uncertainties in the \textit{EE} estimated power spectrum (in red) and the official \textit{Planck} results \cite{PLA_18_I} (in blue). The CENN spectrum corresponds to the results described in Section~\ref{sec:results_realistic_case} and shown in Figure~\ref{Fig:15arcmin_case}. Arrows in the first three multipole bins for the CENN case represent upper limits. Horizontal error bars indicate the binning scheme used in each analysis. The bottom panel shows the difference between the two cases, where the red line represents the ideal case.}
\label{Fig:Uncertainty_EE}
\end{figure}

Furthermore, the comparison between the uncertainty levels in the CENN estimations (in red) and the official \textit{Planck} results \cite{PLA18_I} (in blue) is shown in Figure~\ref{Fig:Uncertainty_EE}. Horizontal bars indicate the multipole bins used in each case. Triangular markers denote upper limits in the CENN estimations, primarily due to the use of small sky patches. Black points represent the difference between the two estimations, while the red horizontal line indicates the ideal case.

The relatively high uncertainty levels in the CENN estimates highlight the need for more large-scale information in the training data to achieve reliable results. In contrast, the \textit{Planck} uncertainties are systematically lower than those of CENN, clearly indicating that, although neural networks can produce compelling CMB maps, further research is required to improve their performance for application to real observations.

For the $B$-mode, consistent with findings in \cite{TRI21}, we observe upper limits across the spectrum, underscoring the necessity of further analysis and methodological improvements in future work.

\subsection{Validation by real observations: Planck PR3 data}
\label{sec:results_planck_observations}

In this section, we evaluate the performance of CENN on the PR3 Planck observations. Importantly, we apply the same trained network used in the previous sections without retraining it. For comparison, we use the inpainted map released by the Planck Collaboration from Commander method (\cite{ERI06}, \cite{ERI08}, \cite{GAL23}, \cite{PLA_18_V}), in which structures along the Galactic plane and bright point sources have been inferred and removed to facilitate component separation.

Consistent with our simulation analyses, we consider the HFI full-mission maps at 100, 143, and 217 GHz, and we aim to output the CMB maps at 143 GHz. All maps, including Commander, are smoothed with a 15' FWHM to improve the signal-to-noise ratio. Sky patches are extracted using the same dimensions as in the validation phase with realistic simulations. To cover the sky efficiently without overlap, we select 100 patches per frequency channel at common sky positions. The Commander map at 143 GHz is also cut at these same positions, enabling a direct comparison. The $Q$ and $U$ patches from both methods are then converted into $E$- and $B$-mode maps using NaMaster.

Given the challenges in recovering $B$-modes from \textit{Planck} data, an issue unresolved by traditional methods, and since our goal here is to conduct a first-order comparison of CENN with established component separation techniques using real observations to see the ability of this kind of methods to generalize from simplistic training data to real observations, we restrict our analysis to the recovery of the $E$-mode power spectrum. A dedicated study of $B$-mode recovery and network optimization for real data is then left for future work.

\begin{figure*}
\centering
\includegraphics[width=15cm, height=2.6cm]{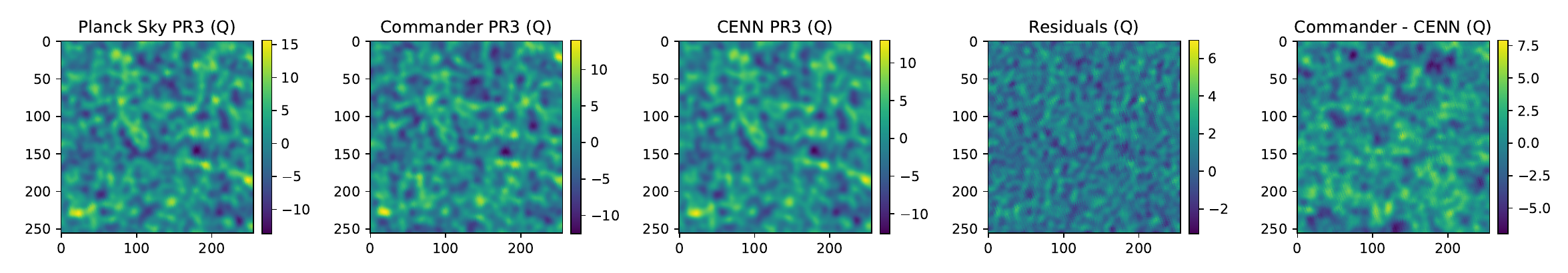}
\includegraphics[width=15cm, height=2.6cm]{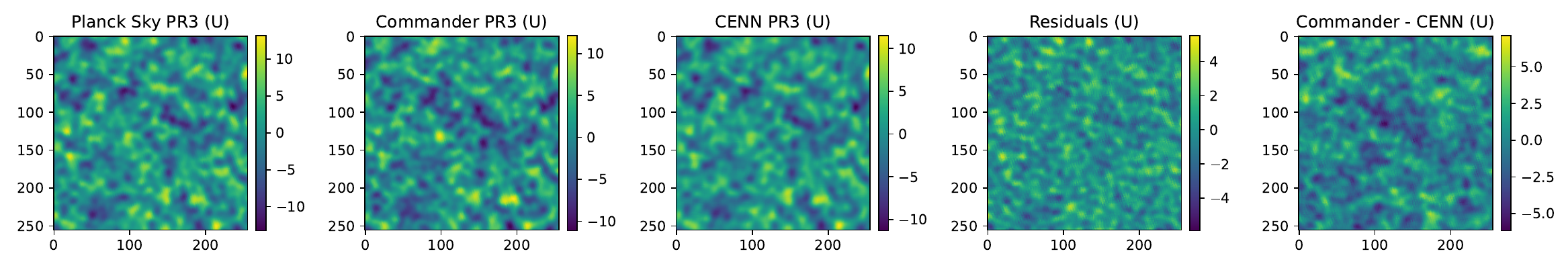}
\caption{Example of the CMB recovery of CENN over one sky patch of Planck PR3 observations (both $Q$ and $U$ maps in top and bottom rows, respectively), smoothed with a 15' FWHM. From left to right: the sky observed by \textit{Planck}, the Commander map, the recovered map by CENN, the residual patch, computed as the difference between both \textit{Planck} observations and CENN output and finally the difference map between Commander and CENN. The units in all the colorbars are $\mu K$.}
\label{Fig:Patches_Observations}
\end{figure*}

Figure \ref{Fig:Patches_Observations} illustrates the CENN output on a representative sky patch from the Planck HFI observations, smoothed at 15' FWHM. The top and bottom rows display the $Q$ and $U$ maps, respectively. The first column shows the original Planck maps, followed (left to right) by the corresponding Commander and CENN outputs, the residual map (difference between Planck and CENN), and finally the difference map between Commander and CENN.

In this example, CENN and Commander exhibit visually broadly consistent behavior. The rightmost column reveals that differences between the two methods are generally confined to a few pixels with extreme values, while the majority of the structures lie within $\pm 2.5~\mu K$. Some residual noise is visible in the CENN output (fourth column), but the residuals remain relatively stable, with mean values around $\pm 2~\mu K$. These residuals could likely be reduced by retraining the network specifically on real data, as was done iteratively for the \textit{Planck} component separation pipelines. The current network architecture serves only as a first-pass model to assess the feasibility of applying deep learning to real CMB observations. Future work will require both architectural refinement and improvements in training dataset quality to achieve optimal results.

\begin{figure}[ht]
\centering
\includegraphics[width=0.7\linewidth]{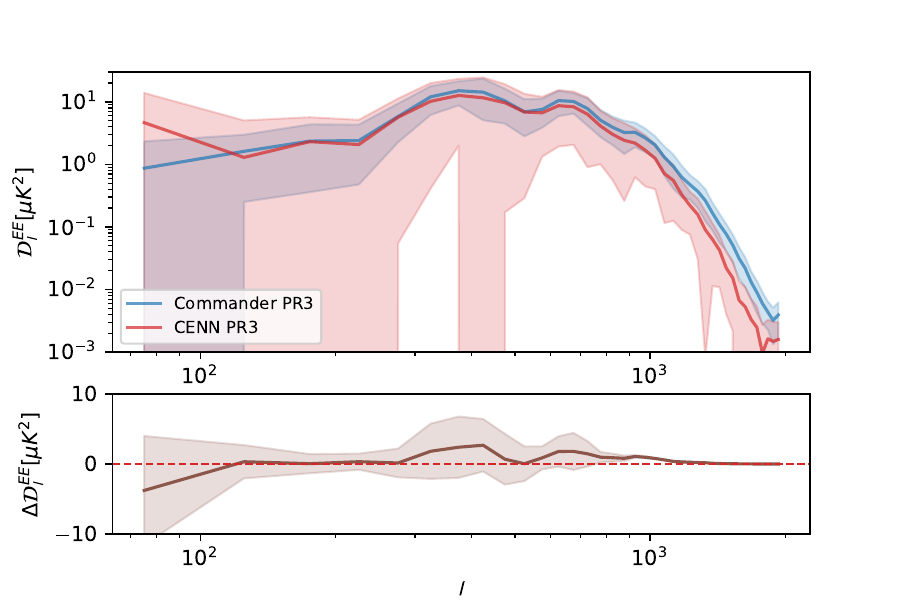}
\caption{Recovered $E$-mode power spectrum at 143 GHz from CENN (in red) applied to the Planck PR3 observations, compared with the Commander PR3 map (in blue).
The bottom subpanel shows the absolute difference between both spectra. Shaded regions indicate the $1\sigma$ uncertainties: for Commander, these are purely statistical; for CENN, they combine statistical and systematic uncertainties.}
\label{Fig:EE_Observations}
\end{figure}

Furthermore, Figure \ref{Fig:EE_Observations} presents the average $EE$ power spectra from the 100 $Q$ and $U$ patches, comparing Commander (in blue) and CENN (in red). The bottom panel shows the difference between the two. Shaded regions indicate $1\sigma$ uncertainties: for Commander, these are statistical (computed from the patch ensemble); for CENN, they include both statistical and systematic contributions, being the latter estimated using realistic simulations.

Overall, the trends are consistent with those from the realistic simulations in Section \ref{sec:results_realistic_case}. Variance is highest at large angular scales, primarily due to the limited patch sizes (we infer significant fluctuations for $l<120$). This is likely compounded by limited large-scale information exposure during training due to the small patch sizes. Notably, the CENN output deviates from Commander’s for $l<300$, where Commander appears more stable likely owing to its all-sky modeling.

Additionally, we observe high statistical uncertainty at $l \sim 400$. Addressing this will require methodological improvements in future work, such as adopting alternative sky projections that permit all-sky coverage. We have not pursued this here due to the risk of border artifacts and the $E$-to-$B$ leakage encountered with larger patches, issues requiring further investigation.

At smaller angular scales, CENN matches Commander within uncertainties. As in Section \ref{sec:results_realistic_case}, we observe a modest underestimation, though the mean error remains below 5\%. This performance could be improved through enhanced training data and network refinements.

Looking ahead, we anticipate in principle improved performance from CENN when applied to current and future observations from next-generation, high-sensitivity experiments such as the Simons Observatory, LiteBIRD, and PICO. Compared to the idealized results in Section \ref{sec:results_ideal_case}, the gains in data quality expected from these missions could enable neural networks like CENN to achieve significantly better component separation. These models should be further developed and evaluated in upcoming forecasting studies \cite{FUS23}. Overall, our results suggest that neural networks seem worth using as methods for component separation in future CMB polarization analyses.

\subsection{Robustness tests}
\label{sec:additional_tests}

To further evaluate the robustness of CENN and identify potential avenues for improvement, we conducted additional tests complementing the main analyses presented in the previous sections.

\begin{figure*}[t]
\centering
\minipage{0.5\textwidth}
\includegraphics[width=\linewidth]{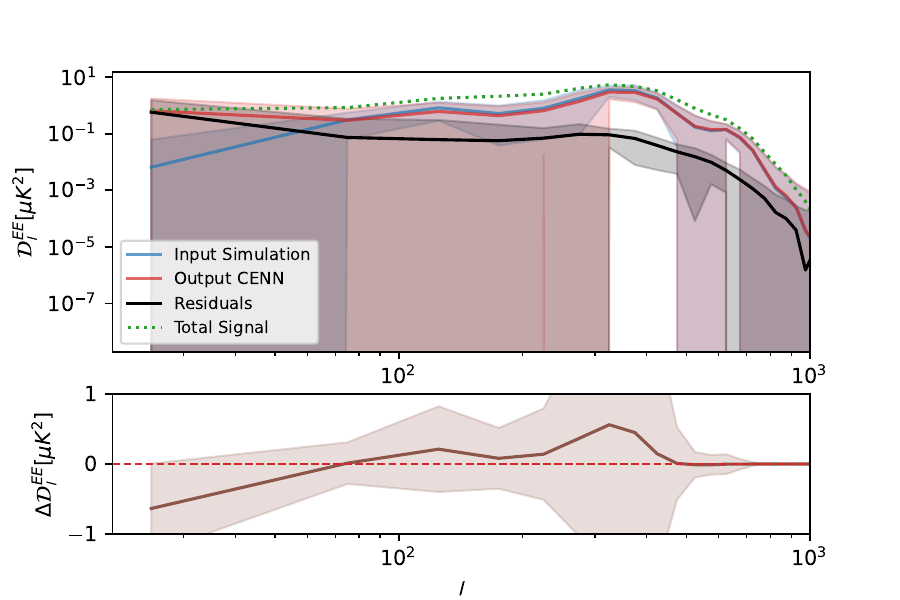}
\endminipage\hfill
\minipage{0.5\textwidth}%
  \includegraphics[width=\linewidth]{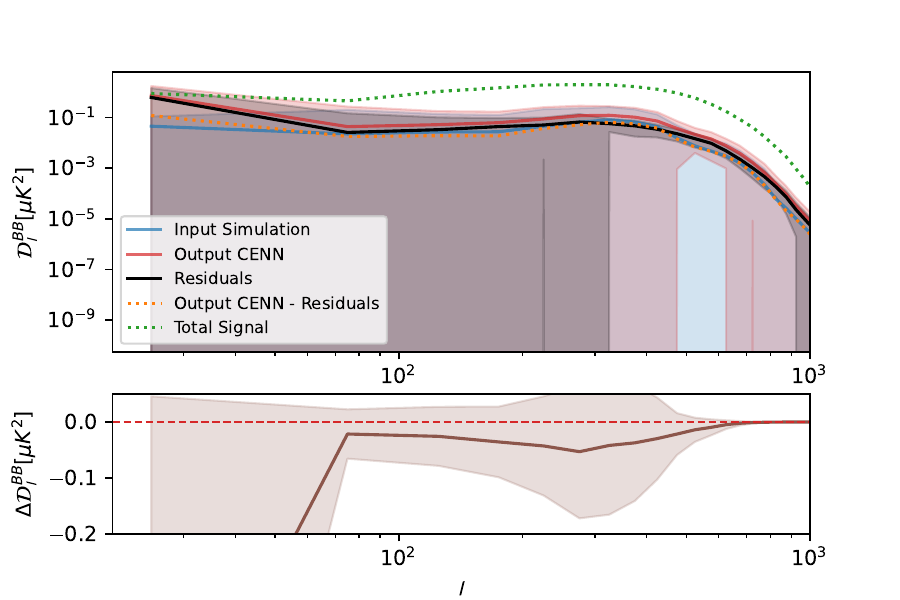}
\endminipage
\caption{$E$- and $B$-mode power spectra recovery by CENN in the realistic case with 30' FWHM smoothing. Top left panel: comparison between the input $E$-mode CMB power spectrum (in blue) and the CENN output (in red); residuals are shown in black. The green dotted line indicates the average total input spectra for the 143 GHz patches. Bottom left panel: absolute difference between the input and recovered spectra. Right panel: same analysis for the $B$-mode; the orange dotted line shows the difference between the output and residual spectra. Shaded areas represent the 1$\sigma$ standard deviation for each multipole bin.}
\label{Fig:30arcmin}
\end{figure*}

Given the well-established role of smoothing in CMB studies for suppressing background noise, our first supplementary test explores the effect of applying a stronger smoothing kernel. Specifically, we validated the network in the same southern sky region ($-60^{\circ} < b < -90^{\circ}$) not used during training, using the realistic simulated data from Section \ref{sec:results_realistic_case} but smoothed with a 30' FWHM kernel. As shown in Figure \ref{Fig:30arcmin}, this increased smoothing yields a modest improvement in the recovered power spectra, with a mean absolute error of $0.3 \pm 0.1 \mu K^{2}$ for the $E$-mode and $-0.05 \pm 0.10 ,\mu K^{2}$ for the $B$-mode, and residuals around $10^{-1} ,\mu K^{2}$ in both cases. However, this small gain comes at the expected cost of losing information at the smallest angular scales. Despite this trade-off, stronger smoothing may be advantageous when the primary scientific goal is to set robust upper limits on the CMB power spectra at selected multipoles (e.g., near the first acoustic peak) rather than to reconstruct the full angular power spectrum, particularly for the challenging $B$-mode signal.

Secondly, we performed a kind of null test to probe the origin of the residual signals and to verify that the CENN output indeed represents an upper bound on the true CMB signal. For this purpose, we applied the network to a validation dataset constructed without any CMB signal, i.e. retaining only foregrounds and instrumental noise, and using the same validation sky region as before. Additionally, to further test the network’s generalization capacity beyond its initial hint on real observations, we validated it using a different foreground model than the one employed during training. Specifically, we used the PySM ‘d4’ and ‘s2’ dust and synchrotron models, which produce a physically distinct sky configuration from the PLA models used in training\footnote{For details on the PySM models and their differences from the PLA templates, see the PySM3 documentation: https://pysm3.readthedocs.io/en/latest/models.html.}.

\begin{figure*}[t]
\centering
\minipage{0.5\textwidth}
\includegraphics[width=\linewidth]{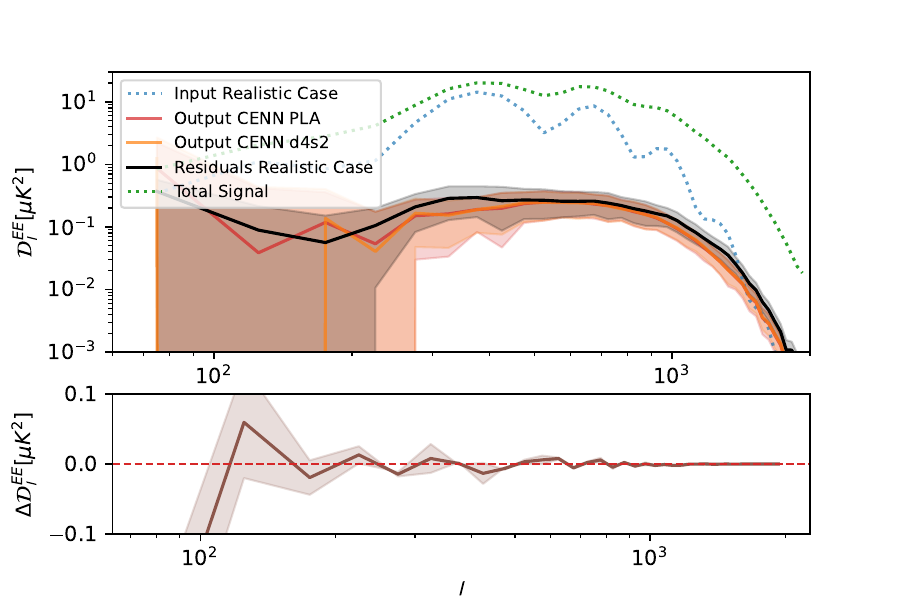}
\endminipage\hfill
\minipage{0.5\textwidth}%
  \includegraphics[width=\linewidth]{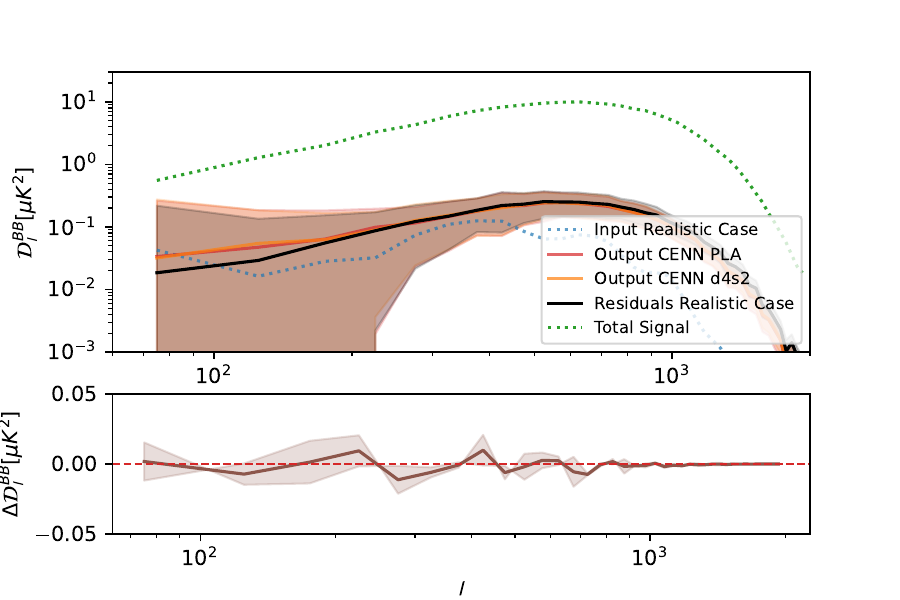}
\endminipage
\caption{$E$- and $B$-mode power spectra recovery by CENN in a realistic null-test scenario (without an input CMB signal) at 15' FWHM smoothing. Top left panel: comparison between the input $E$-mode CMB spectrum from the 15' case in Section \ref{sec:results_realistic_case} (blue dotted line) and the CENN outputs when validated with realistic PLA simulations (in red) and PySM d4s2 foreground models (in orange). Residuals from the 15' realistic case are shown in black. The green dotted line indicates the average Input Total spectra at 143 GHz. Bottom left panel: absolute difference between the hypothetical input and output spectra. Right panel: same analysis for the $B$-mode. Shaded areas show the 1$\sigma$ standard deviation for each bin.}
\label{Fig:null_test}
\end{figure*}

Figure \ref{Fig:null_test} illustrates the CENN outputs for the $E$- and $B$-mode spectra (left and right panels, respectively) recovered from skies generated with the PLA models (in red) and the alternative PySM model (in orange). The respective subpanels show the difference between the two. For reference, the input and residual signals from the realistic case in Section \ref{sec:results_realistic_case} are shown in blue and black, and the Input Total signal at 143 GHz in green.

Overall, the outputs for both sky realizations show similar shapes and are generally lower in amplitude than the residuals found in the realistic case, especially for the $E$-mode. This result has two important implications: first, it confirms that a significant portion of the residuals in the realistic scenario originates from instrumental noise, which the network partially misinterprets as a true CMB signal, regardless of the specific foreground sky used. Second, and crucially for observational cosmology, this kind of null test demonstrates that CENN does not generate spurious CMB power exceeding the level recovered when a true CMB signal is present. This property is essential for using CENN outputs as conservative upper bounds on the CMB power spectra, particularly for the faint $B$-mode when applied to real observations.

Furthermore, when validating the network on an alternative sky model, the recovered spectra remain consistent, showing only small deviations at large angular scales (typically 5–10\% for both modes). While more extensive tests with diverse foreground configurations are necessary for future applications, this initial result indicates the network’s promising capacity to generalize to different microwave sky conditions at relevant angular scales. Finally, we reiterate a key limitation: due to the current patch size, our analysis does not yet robustly probe the largest angular scales, highlighting an area for future methodological development.

\section{Conclusions}
\label{sec:conclusions}

In this work, we have extended the application of the Cosmic Microwave Background Extraction Neural Network (CENN), previously shown to be effective for temperature component separation, to the more challenging task of recovering the polarized CMB signal.

CENN is a convolutional neural network that processes three input sky patches at 100, 143, and 217 GHz, as observed by \textit{Planck}, and is trained to reconstruct the CMB $Q$ and $U$ Stokes parameter maps at 143 GHz. The training set consists of realistic simulations including CMB, Galactic and extragalactic foregrounds, and instrumental noise, with patches extracted from the region $90^{\circ} < b < -60^{\circ}$. Validation is performed on an independent region ($-60^{\circ} < b < -90^{\circ}$) not seen during training. The network's output is evaluated by comparing the recovered $EE$ and $BB$ power spectra and the residual spectra (the power spectrum of the difference between input and output CMB maps).

In an ideal, noise-free scenario, CENN demonstrates promising performance to separate the polarized CMB signal from foregrounds, achieving mean absolute errors for $l>200$ of approximately $0.5 \pm 1~\mu K^{2}$ for the $E$-mode and $0.05 \pm 0.1~\mu K^{2}$ for the $B$-mode, corresponding to relative errors below 5\% for $E$ and about 10\% for the more challenging $B$-mode, with residuals in the range $10^{-2}$–$10^{-1}~\mu K^{2}$. Although higher variance is observed in the $B$-mode, especially at large angular scales, these results indicate that, with methodological improvements such as enhanced projection techniques, neural networks like CENN may be useful for future high-sensitivity CMB experiments. Extending this work to forecasting for missions like the Simons Observatory, LiteBIRD, and PICO remains an important direction for future research.

In a more realistic scenario including \textit{Planck}-level isotropic white instrumental noise, residual power increases to about $10^{-1}~\mu K^{2}$ for both modes, with a mean absolute error of approximately $0.1 \pm 0.3~\mu K^{2}$ for $E$ and $-0.1 \pm 0.3~\mu K^{2}$ for $B$, confirming that the added noise is the primary contributor to this higher residual level. Nonetheless, the $E$-mode is recovered with moderate fidelity, while the $B$-mode remains strongly affected by noise. These findings suggest that performance improvements with lower-noise data from next-generation experiments are possible but remain to be demonstrated.

However, for these kinds of methods to be employed in future surveys targeting high-precision cosmological estimations and reliable CMB maps and power spectrum reconstructions, significant improvements in the quality of training data are necessary to reduce the current high uncertainties associated with the methodology.

Although this work provides a first approximation of the performance of such methods when applied to this type of data and analysis, further development is required. Future efforts should incorporate more realistic sky signals and complex instrumental noise models, including scan-path-modulated white noise and/or full end-to-end simulations.

Furthermore, at large angular scales, residuals remain comparable to the ideal case, with only minor foreground contamination below $\ell \sim 80$. This indicates that, within the current framework, limitations at large scales arise mainly from the finite patch size rather than instrumental noise, highlighting an area for future methodological refinement.

Residual analysis in the realistic case shows that although the $B$-mode is more sensitive to noise and foregrounds, the CENN output retains a conservative representation of the true CMB signal, even when the signal is weaker than the noise. Attempts to further clean the output using a secondary neural network produced similar results, implying that with the present architecture and Planck-like noise levels, the residuals approach the statistical limit for CMB recovery, especially for the faint $B$-modes.

Applying the trained network to the Planck PR3 data, using 100 sky patches at the same frequencies, we find that the estimated $EE$ power spectra agree with those from the Commander maps within 5\% at intermediate and small scales, with slight differences at high multipoles, although higher uncertainty levels with respect Commander maps have been found in our analysis. We suggest that improvements in the training data and architecture should also reveal better performance in future works with trainings especially dedicated to the application of these methods in real data. Also, larger scales remain dominated by statistical uncertainty, reinforcing the need for improved patch-based methodologies for future satellite missions.

Additional tests with stronger smoothing (30' FWHM) yielded slight improvements in the recovery capacity but at the expense of angular resolution. This suggests that stronger smoothing may be beneficial for setting conservative upper limits near specific multipoles, such as the first acoustic peak.

Finally, a robust null test applying CENN to simulations without an input CMB signal and validating on alternative foreground models confirmed that residual signals are mainly due to instrumental noise, with no spurious CMB recovery beyond the level found when the true CMB is present. The network’s performance was consistent across different foreground realizations, with only minor large-scale differences.

In summary, these results demonstrate that neural network methods like CENN appear to be promising tools for component separation in CMB polarization studies for future experiments. However, improvements in both the methodology and the training data are necessary before these methods can be reliably applied to real observations and high-precision cosmological analyses. Current limitations have been identified, particularly in low signal-to-noise regimes and at large angular scales which highlight the need for further methodological advances, including future training directly on the sphere.

\acknowledgments
The authors would like to warmly thank the editor for their comments that highly improved this paper. JMC acknowledges the PID2023-151567NB-I00 project. JMC, LB and JGN acknowledge the CNS2022-135748 project. JMC, LB, JGN and DC acknowledge the PID2021-125630NB-I00 project. JMC and SRC also acknowledge the SV-PA-21-AYUD/2021/51301 grant. GP acknowledges support from Italian Research Center on High Performance Computing Big Data and Quantum Computing (ICSC), project funded by European Union NextGenerationEU and National Recovery and Resilience Plan (NRRP) Mission 4 Component 2 within the activities of Spoke 3 (Astrophysics and Cosmos Observations). CB acknowledges support from the COSMOS project of the Italian Space Agency, and the INDARK Initiative of the INFN. CGC and FJDC acknowledge the PID2021-127331NB-I00 project.\\
This research has made use of the packages \texttt{Matplotlib} \citep{matplotlib}, \texttt{Keras} \citep{KER}, \texttt{Numpy} \citep{numpy}, \texttt{Namaster} \citep{NAMASTER}, \texttt{HEALPix} \citep{GOR05} and \texttt{Healpy} \citep{zon19} packages.

\bibliographystyle{JHEP}
\bibliography{main}

\end{document}